# How are journals cited? characterizing journal citations by type of citation

Domenic Rosati[1]

[1]*dom@scite.ai*
scite, Inc., Brooklyn, NY, USA

## Introduction

Evaluation of journals for quality, impact, and prestige is one of the dominant themes of bibliometrics since journals are the primary venue of vetting and distribution of scholarship. There are many criticisms of quantifying journal impact with bibliometrics including disciplinary differences among journals, what source materials are used, time windows for the inclusion of works to measure, and skewness of citation distributions (Lariviere & Sugimoto, 2019). However, despite various attempts to remediate these in newly proposed indicators, such as SJR, SNIP, and Eigenfactor, journal evaluation still remains based on citation counts and fails to acknowledge the critical differences that the type of citation makes (Walters, 2017). That is whether the citation is supporting or disputing a work when quantifying journal impact and quality.

Various initiatives have been suggested to encompass citation content analysis within bibliometrics projects (Shotton 2010; Kilicoglu 2019; Xu et. al. 2015). However, citation content analysis has not been done at the scale needed in order to perform meaningful journal analysis since that would require the classification of hundreds of millions of citations based on their function such that those classifications could be counted at parity with what is available in standard citation indexes such as Scopus and Web of Science. The scite smart citation index is now available with citation classifications at the scale needed to compare journals with traditional bibliometrics tools (scite). scite classifies citations based on the rhetorical function of a citation: whether the citation supports, disputes, or simply mentions another article. Currently, scite has classified 823 million citation statements and we are able to use this data at scale to look meaningfully at the distribution of supporting and disputing citations received by journals and contrast those indicators with previous work in bibliometrics.

## Methods

In order to look at how journals receive supporting and disputing citations, we have aggregated citations type counts as tallies for individual journals. Using standard techniques of descriptive statistics, we are able to look at how supporting and disputing citations are characterized. We are also able to present how supports and disputes are correlated with citation counts, using the Pearson correlation coefficient, and what the distribution of these citation types look like at scale.

In addition to characterizing counts of citation types for journals, we also wanted to ask what the ratio of supporting and disputing citations looks like. That is out of all supporting and disputing citations received how many of those are supporting? We hypothesize that this presents a normalized indicator that captures a notion of journal quality since it would indicate the overall ratio of how well-supported publications from a journal are. We call this metric the "scite index" (SI) of a journal and calculate it as:

$$SI = \frac{\text{\# Supporting Cites}}{\text{\# Supporting Cites} + \text{\# Disputing Cites}}$$

In order to make this ratio more informative, we only calculate the index for journals with more than 100 citations received and at least 1 citation classification made. We calculate this index based on all the citations available from scite without a time window.

## Results

Table 1 presents the descriptive characteristics of supports and disputes received by a journal as well as its scite index. This shows that supporting and disputing citations received by a journal display the characteristic heavy skew that is common in the statistics of citation counts and bibliometrics for journals based on those counts. This result seems intuitive since supporting and disputing citations are just a proportion of the total citations which are known to have these characteristics. This is well established in Table 2 which presents a very strong correlation of supporting citations to a journal with its total citations and the same with its disputing citations.

With the scite index, we found two surprising results. The first is that the correlations presented in Table 2 show that there is no correlation between citation count and the scite index (presented graphically as a in Figure 1). The second is that the distribution of this ratio is not skewed, showing normal distribution of well-supported journals (presented in Figure 2).

**Table 1. Descriptive characteristics of supporting, disputing, citations received by a journal and their scite index**

| Table | Supporting | Disputing | Scite Index |
|---|---|---|---|
| Journal Count | 65313 | 65313 | 30929 |
| Mean | 513 | 64 | 0.87 |
| Median | 2 | 0 | 0.88 |
| SD | 5434 | 608 | 0.08 |
| Skew | 46 | 39 | -2.14 |
| Min | 0 | 0 | 0.25 |
| Max | 531746 | 55183 | 0.99 |

**Table 2. Pearson correlation coefficient for citation types and scite index with total citations received per journal..**

| Table | Supporting | Disputing | scite index |
|---|---|---|---|
| Total Citations | R=0.96 | R=0.96 | R=0.04 |

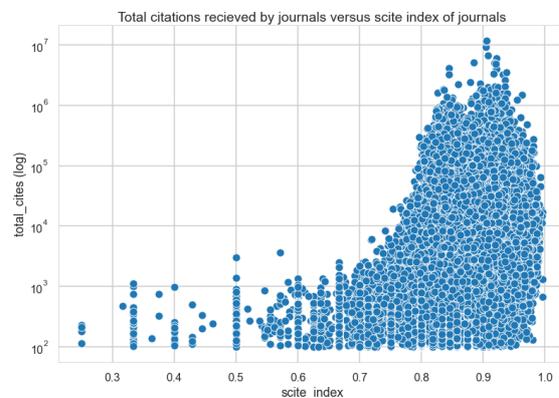

**Figure 1. Total citations received by journal versus scite index of journals (total_cites are presented on a log scale).**

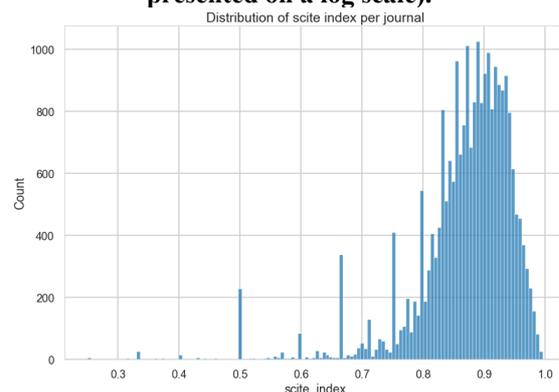

**Figure 2. Distribution of scite index per journal**

**Discussion**

What our findings suggest is that while support and dispute counts are unsurprisingly similar to citation counts in general, if we use the ratio of supported publications to supported and disputed ones as an indicator of journal quality, we find that well-supported journals are normally distributed. The importance of this finding is that journal indicators tend to be heavily skewed reinforcing the fact that heavily cited and well-established journals tend to dominate indicators of impact, prestige, and quality (Lariviere & Sugimoto, 2019). If we take the scite index of journals as an indicator of quality based on how well supported the publications in the journal are, then perhaps this indicator provides a more equitable approach to evaluating journal quality. In order to evaluate the full impact of this, we are working on correlating journal impact metrics with citation type data in a future study to explore whether current journal metrics capture the information provided by citation types. In addition, future studies should look at other aggregate levels such as authors and institutions or look at the distribution of support and dispute among individual articles in order to provide nuance to research in bibliometrics on those levels as well. It is our hope that these preliminary findings provide a potential window in how to add nuance to journal evaluation by adding additional information from citation content analysis.

**Data availability**
Data and code available on request

**Acknowledgements**
Thanks to the scite team for providing these citation data, supporting this research, and feedback.

**References**
Kilicoglu, H., Peng, Z., Tafreshi, S., Tran, T., Rosemblat, G., & Schneider, J. (2019). Confirm or refute?: A comparative study on citation sentiment classification in clinical research publications. Journal of biomedical informatics, 91, 103123.
Lariviere, V., & Sugimoto, C. R. (2019). The journal impact factor: A brief history, critique, and discussion of adverse effects. In Springer handbook of science and technology indicators (pp. 3-24). Springer, Cham.
scite (n.d.). Retrieved February 21, 2021, from https://scite.ai/
Shotton, D. (2010, June). CiTO, the citation typing ontology. In Journal of biomedical semantics (Vol. 1, No. 1, pp. 1-18). BioMed Central.
Walters, W. H. (2017). Citation-based journal rankings: Key questions, metrics, and data sources. IEEE Access, 5, 22036-22053.
Xu, J., Zhang, Y., Wu, Y., Wang, J., Dong, X., & Xu, H. (2015). Citation sentiment analysis in clinical trial papers. In AMIA annual symposium proceedings (Vol. 2015, p. 1334). American Medical Informatics Association.